# SPEAKER RECOGNITION IN BENGALI LANGUAGE FROM NONLINEAR FEATURES

Uddalok Sarkar[1], Soumyadeep Pal[1], Sayan Nag[1], Chirayata Bhattacharya[4],
Shankha Sanyal[2,3]*, Archi Banerjee[2,3], Ranjan Sengupta[2] and Dipak Ghosh[2]
[1] Department of Electrical Engineering, [2] Sir C.V. Raman Centre for Physics and Music,
[3] Department of Physics, [4] Department of Electronics & Telecommunication Engineering
Jadavpur University
* Corresponding Author

**ABSTRACT**

*At present Automatic Speaker Recognition system is a very important issue due to its diverse applications. Hence, it becomes absolutely necessary to obtain models that take into consideration the speaking style of a person, vocal tract information, timbral qualities of his voice and other congenital information regarding his voice. The study of Bengali speech recognition and speaker identification is scarce in the literature. Hence the need arises for involving Bengali subjects in modelling our speaker identification engine. In this work, we have extracted some acoustic features of speech using non linear multifractal analysis. The Multifractal Detrended Fluctuation Analysis reveals essentially the complexity associated with the speech signals taken. The source characteristics have been quantified with the help of different techniques like Correlation Matrix, skewness of MFDFA spectrum etc. The Results obtained from this study gives a good recognition rate for Bengali Speakers.*

## INTRODUCTION

Automatic Speaker recognition refers to identification of "who is speaking" on the basis of features of their respective speech waves. There are two major aspects of Automatic speaker recognition, speaker verification and speaker identification [1]. In Identification task the system has to identify the speaker (class) from a set of known voices (training dataset); whereas the speaker verification task involves accepting or rejecting identity claim of a speaker, where Speaker can be from open set or closed set. ASR can also be classified into two type: text dependent Automatic Speaker Recognition, text independent Automatic Speaker Recognition. In text dependent ASR the same phrase or key word is used for both enrollment and verification. Hence, the system knows the phrase to be uttered by the speaker beforehand. This type of system is much accurate. But text independent ASR is much robust and flexible but hard to design and less accurate, as the utterance of the speakers is different in enrollment and verification [2].

Most of the previous studies on Automatic Speaker recognition uses statistical characteristics related to temporal, spectral and cepstral features of each frame of the whole speech wave. Some temporal features for speaker identification are Zero crossing rate, Signal Energy, Maximum Amplitude and some useful spectral features are pitch contour, spectral Centroid, spectral flux, Perceptive Linear Prediction (PLP). Mel frequency Cepstral Coefficients is most frequently used cepstral feature in the field of Automatic Speaker Recognition [3]. Hence, most of the popular features that are used for speaker identification are primarily based on temporal to frequency domain transformation of speech wave. All these frequency domain techniques primarily use Fourier Transformation for transforming the signal into frequency domain. This method is strongly questioned for non-stationary aspect of signal. Numerous high frequency harmonics are left unattended in Fourier spectral analysis [4].

Non-linear dynamical modeling for source clearly indicates the relevance of non-deterministic /chaotic approaches in understanding the speech signals [5-7]. In this context fractal analysis of the speech signal which reveals the geometry embedded in signal assumes significance. Fractal analysis of audio signals was first performed by Voss and Clarke [8], who analyzed amplitude spectra of audio signals to find out a characteristic frequency f c , which separates white noise (which is a measure of flatness of the power spectrum) at frequencies much lower than f c from very correlated behavior (1/f$^2$ ) at frequencies much higher than f c . Speech data is essentially a quantitative record of variations of a particular quality over a period of time. However, it is well-established experience that naturally evolving geometries and phenomena are rarely characterized by a single scaling ratio; different parts of a system may be scaling differently. That is, the clustering pattern is not uniform over the whole system. Such a system is better characterized as 'multifractal' [9]. A multifractal can be loosely thought of as an interwoven set constructed from sub-sets with different local fractal dimensions. Real world systems are mostly multifractal in nature. Speech too, has non-uniform property in its movement [10,11]. In a number of recent studies [11, 12], Multifractal Detrended Fluctuation Analysis (MFDFA) [9]have been applied to extract specific features of different music clips.

In this study, for the first time, we have applied MFDFA on the Bengali speech corpus of 5 participants to extract specific features which can help in development of an efficient speaker recognition algorithm in respect to Bengali speech. The multifractal spectral width generated from the self-similar speech signals have been used as a parameter to extract robust features from the different speech signals recorded from the speakers. An attempt has been made here to model a Multi-fractal Detrended Fluctuation Analysis based text independent Automatic Speaker recognition system. The study reveals new and interesting information which can pave the way for a number of new avenues in the domain of nonlinear approach to automatic speaker recognition. The same can be further expanded in regard to different languages.

**EXPERIMENTAL DETAILS**

Five (5) paragraphs selected from widely popular Bengali texts by eminent novelists were taken for analysis. The recordings were done from 5 (five) speakers who were asked to read the text normally without any emotional content being imbibed in them. The signals are digitized at the rate of 22050 samples/sec 16 bit format. The readings of each script have been kept within time duration of more or less 2 minutes and similar phrases (which were about 10 seconds duration) have been extracted then after for the ease of analysis. The fractal analysis of different segments has been carried out separately to get the necessary multifractal measures.

**METHOD OF ANALYSIS**
**Method of multifractal analysis of sound signals**
The time series data obtained from the sound signals are analyzed using MATLAB [13] and for each step an equivalent mathematical representation is given which is taken from the prescription of Kantelhardt et al [9].
The complete procedure is divided into the following steps:
*Step 1:* Converting the noise like structure of the signal into a random walk like signal. It can be represented as:

$$Y(i) = \sum (x_k - \bar{x}) \quad (1)$$

Where $\bar{x}$ is the mean value of the signal.
*Step 2:* The local RMS variation for any sample size *s* is the function *F(s,v)*. This function can be written as follows:

$$F^2(s,v) = \frac{1}{s}\sum_{i=1}^{s} \{Y[(v-1)s+i] - y_v(i)\}^2 \quad (2)$$

*Step 4:* The q-order overall RMS variation for various scale sizes can be obtained by the use of following equation

$$F_q(s) = \left\{ \frac{1}{Ns} \sum_{v=1}^{Ns} [F^2(s,v)]^{\frac{q}{2}} \right\}^{\left(\frac{1}{q}\right)} \qquad (3)$$

*Step 5:* The scaling behaviour of the fluctuation function is obtained by drawing the log-log plot of $F_q(s)$ vs. s for each value of q.

$$F_q(s) \sim s^{h(q)} \qquad (4)$$

The h(q) is called the generalized Hurst exponent. The Hurst exponent is measure of self-similarity and correlation properties of time series produced by fractal. The presence or absence of long range correlation can be determined using Hurst exponent. A monofractal time series is characterized by unique h(q) for all values of q.

The generalized Hurst exponent h(q) of MFDFA is related to the classical scaling exponent τ(q) by the relation

$$\tau(q) = qh(q) - 1 \qquad (5)$$

A monofractal series with long range correlation is characterized by linearly dependent q order exponent τ(q) with a single Hurst exponent H. Multifractal signal on the other hand, possess multiple Hurst exponent and in this case, τ(q) depends non-linearly on q [9].

The singularity spectrum f(α) is related to h(q) by

$$\alpha = h(q) + qh'(q) \qquad (6)$$
$$f(\alpha) = q[\alpha - h(q)] + 1 \qquad (7)$$

Where α denoting the singularity strength and *f(α)*, the dimension of subset series that is characterized by α. The width of the multifractal spectrum essentially denotes the range of exponents. The spectra can be characterized quantitatively by fitting a quadratic function with the help of least square method [9] in the neighbourhood of maximum $\alpha_0$,

$$f(\alpha) = A(\alpha - \alpha_0)^2 + B(\alpha - \alpha_0) + C \qquad (8)$$

Here C is an additive constant C = f(α0) = 1 and B is a measure of asymmetry of the spectrum. So obviously it is zero for a perfectly symmetric spectrum. We can obtain the width of the spectrum very easily by extrapolating the fitted quadratic curve to zero.
Width W is defined as,

$$W = \alpha_1 - \alpha_2 \qquad (9)$$

with

$$f(\alpha_1) = f(\alpha_2) = 0$$

The width of the spectrum gives a measure of the multifractality of the spectrum. Greater is the value of the width W greater will be the multifractality of the spectrum. For a monofractal time series, the width will be zero as h(q) is independent of q. The spectral width has been considered as a parameter to evaluate how the features of speech of a particular speaker varies from another. Different features have been extracted using the multifractal curve as the input parameter.

**Statistical Features of MFDFA Spectrum:**

To extract statistical features from the MFDFA spectrum we had to normalize the singularity spectrum $f(\alpha)$ by,-

$$f_n(\alpha) = \frac{f(\alpha)}{\sum_\alpha f(\alpha)} \qquad (10)$$

Then we treated MFDFA spectrum as with normalized singularity spectrum as a probability distribution. We have taken two statistical features of this distribution for classification purpose. These are:
1. <u>Feature-1: difference between median and mode-</u> The mathematical expression we used to determine this value is,-
$$Feature1 = \text{median}(\alpha) - \underset{\alpha}{\arg\max}\, f_n(\alpha) \qquad (11)$$

2. <u>Feature-2: Skewness</u> – Mathematical Expression used for Skewness is,-

$$Feature2 = E\left(\frac{\alpha - \mu}{\sigma}\right) \quad (12)$$

Where, normalized $f_n(\alpha)$ is treated as pmf of discrete random variable $\alpha$.

## RESULTS AND DISCUSSION

The following figure (**Fig. 1**) is a representative display of the multifractal spectrum obtained for five different speakers

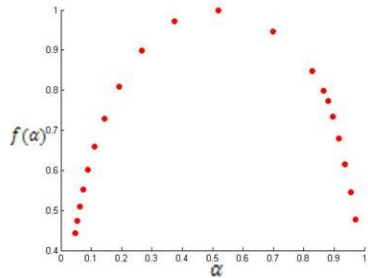
Fig. 1 (a)

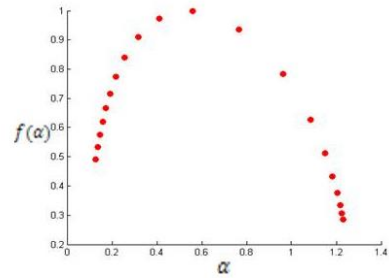
Fig. 1 (b)

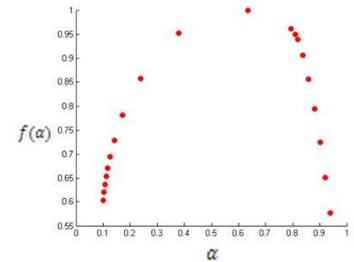
Fig. 1 (c)

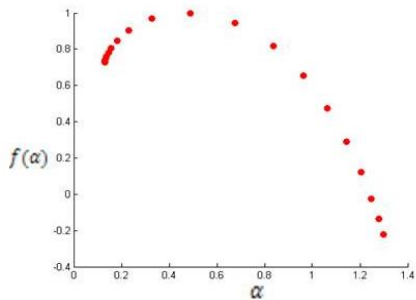
Fig. 1 (d)

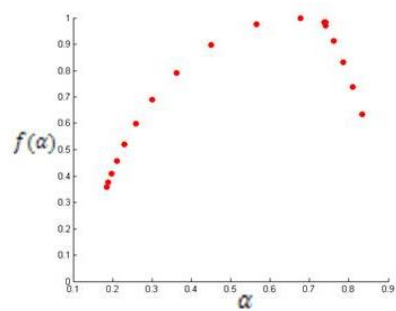
Fig. 1 (e)

**Fig. 1:** MFDFA spectrum instances- a)Speaker1  b) Speaker2  c) Speaker3  d) Speaker4  e) Speaker5

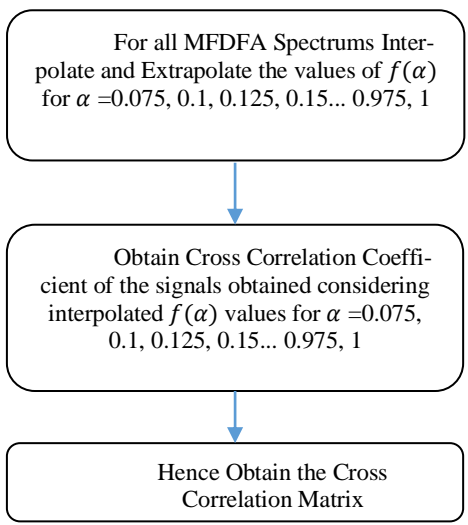

**Fig. 2**: Protocol for determining cross correlation coefficient of MFDFA spectrum

From Fig.1 We have an intuition of how the MFDFA spectrum varies from speaker to speaker. To emphasis more on this fact we tried to extract cross correlation coefficients between MFDFA spectrums of different speech waves. To obtain these cross correlation coefficients we followed protocol of Fig.2.

| Speech Index in Correlation Matrix | Speaker Identity |
|---|---|
| 0 to 20 | Speaker1 |
| 21 to 40 | Speaker2 |
| 41 to 60 | Speaker3 |
| 61 to 80 | Speaker4 |
| 81 to 100 | Speaker5 |

**Table 1.** Speech data index versus speaker Identity

In our Experiment we had 5 subject speakers and 20 Speech waves of each Speaker. Hence we came up with 100 total speech waves and hence a Cross Correlation Matrix of size 100x100. In *Table1* the Speech Data and its corresponding speaker information are tabulated. In Fig.3 the Correlation matrix is shown which is of size 100x100.

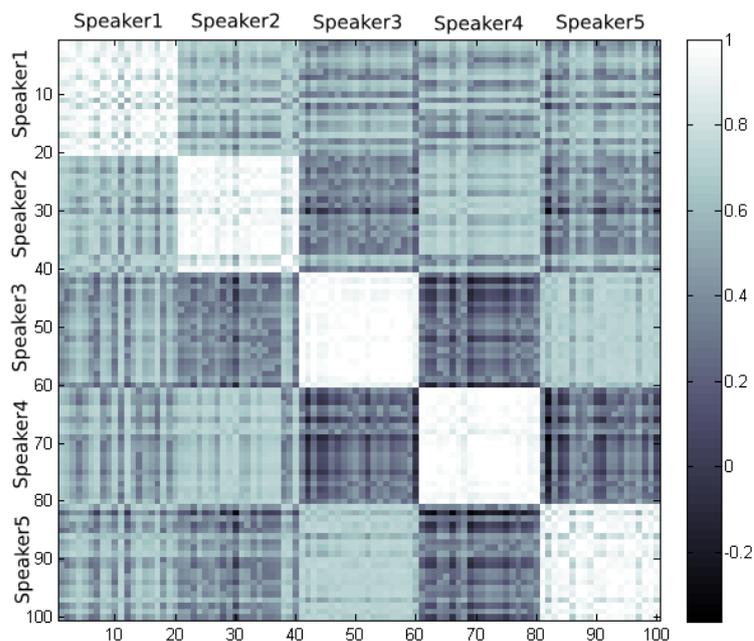

**Fig. 3**: Correlation Matrix

In Fig. 3 the Correlation matrix clearly shows that the Cross Correlation coefficients between MFDFA spectrums of speech waves in case of same speakers are very high but for different speakers these coefficients are very low. For instance, the Correlation Matrix shows that MFDFA spectrum for speaker 4 and speaker 3 are very much dissimilar (having correlation coefficient in the range -0.2 - 0.2), almost the same response can be seen in case of speaker 4 and speaker 5. But, in case of speaker 2 and speaker 4, we find that the values of cross-correlation coefficient are higher than the previous case, i.e. in the range 0.4-0.6. Hence we can safely assume that there exist certain similarities in the spectra of speaker 2 and 4, as well as in Speaker 3 and 5. The feature plot of Fig. 4 gives a complete validation in support to the inferences drawn from this correlation matrix.

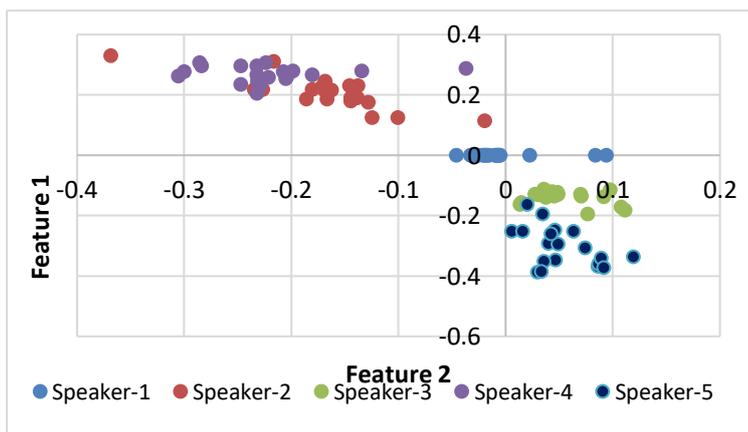

**Fig 4.** Position of Speakers on the 2-dimensional feature space

**Fig. 4** shows a feature plot where the 5 speakers have been shown as a scatter diagram in 2-D feature space. In this plot, the X-direction is Feature 1 and Y-direction is Feature 2.

We can infer from Fig. 4 that for speaker2 and speaker4 the MFDFA spectrum is skewed in same direction with higher Skewness measure in case of speaker4. And for speaker2 and speaker4 $\operatorname*{argmax}_{\alpha} f_n(\alpha) < median(\alpha)$. Similarly, for speaker3 and speaker5 the spectrum is skewed in same direction and speaker4 $\operatorname*{argmax}_{\alpha} f_n(\alpha) > median(\alpha)$. For speaker1 the Skewness measure is almost zero and $\operatorname*{argmax}_{\alpha} f_n(\alpha) = median(\alpha)$.

We have seen the very same instances in Fig.1(a), Fig.1(b), Fig.1(c), Fig.1(d) and Fig.1(f).

After extracting these features for the whole speech database we fitted it with support vector machine. Using RBF kernel for SVM and with a hold-out ratio of ¼ we got the overall accuracy of 96% for speaker identification.

**Table 2.**

Confusion Matrix of speaker identification

| | | Classifier Prediction | | | | | |
|---|---|---|---|---|---|---|---|
| | | Speaker1 | Speaker2 | Speaker3 | Speaker4 | Speaker5 | Total |
| Actual Value | Speaker1 | 5 | 0 | 0 | 0 | 0 | 100% |
| | Speaker2 | 0 | 4 | 0 | 0 | 0 | 100% |
| | Speaker3 | 0 | 0 | 5 | 0 | 0 | 100% |
| | Speaker4 | 0 | 1 | 0 | 5 | 0 | 83.3% |
| | Speaker5 | 0 | 0 | 0 | 0 | 5 | 100% |
| | Total | 100% | 80% | 100% | 100% | 100% | 96% |

In Table 2 the confusion matrix between classifier prediction and true classes is shown. Prediction accuracy of speaker 2 is seen to be 80%. one of the speaker 2 instances is seen to be predicted as speaker 4 instance. In most of the cases, it is seen that the prediction accuracy is 100% using this technique, which makes it a effective and robust method for speaker recognition in the nonlinear scenario.

**CONCLUSION**

Speaker identification through different acoustic features has been the domain of extensive research in the field of speech signal processing. Till date, most of the studies dealt with linear features which characterize the speaker. In this work, for the first time robust nonlinear techniques have been applied to characterize the speech samples corresponding to Bengali language from 5 speakers. Several features like correlation matrix, skewness have been extracted from the nonlinear multifractal spectrum to classify the speakers. The findings have been put through SVM classifier which resulted in 96% classification accuracy in the features applied. This pilot study has immense potential to be applied in different languages to provide a robust algorithm for effective speaker recognition.

**REFERENCES**

[1] Zhu, Qifeng, and Abeer Alwan. "Non-linear feature extraction for robust speech recognition in stationary and non-stationary noise." *Computer speech & language* 17.4 (2003): 381-402.
[2] Chen, Wen-Shiung, and Jr-Feng Huang. "Speaker recognition using spectral dimension features." *Computing in the Global Information Technology, 2009. ICCGI'09. Fourth International Multi-Conference on*. IEEE, 2009.
[3] Md Afzal Hossan Thesis on "Automatic Speaker Recognition Dynamic Feature Identification and Classification using Distributed Discrete Cosine Transform Based Mel Frequency Cepstral Coefficients and Fuzzy Vector Quantization", RMIT
[4] Bhaduri, S., and D. Ghosh. "Non-invasive detection of alzheimer's disease—multifractality of emotional speech." *J. Neurol. Neurosci* (2016).
[5] Behrman, A. (1999). Global and local dimensions of vocal dynamics. *JOURNAL-ACOUSTICAL SOCIETY OF AMERICA*, *105*, 432-443.
[6] Kumar, A., & Mullick, S. K. (1996). Nonlinear dynamical analysis of speech. *The Journal of the Acoustical Society of America*, *100*(1), 615-629.
[7] Sengupta, R., Dey, N., Nag, D., & Datta, A. K. (2001). Comparative study of fractal behavior in quasi-random and quasi-periodic speech wave map. *Fractals*, *9*(04), 403-414.
[8] Voss, R. F., and J. Clarke. "1/f noise in speech and music." *Nature* 258 (1975): 317-318.
[9] Kantelhardt, J. W., Zschiegner, S. A., Koscielny-Bunde, E., Havlin, S., Bunde, A., & Stanley, H. E. (2002). Multifractal detrended fluctuation analysis of nonstationary time series. *Physica A: Statistical Mechanics and its Applications*, *316*(1-4), 87-114.
[10] Lopes, R., & Betrouni, N. (2009). Fractal and multifractal analysis: a review. *Medical image analysis*, *13*(4), 634-649.
[11] Sanyal, S., Banerjee, A., Patranabis, A., Banerjee, K., Sengupta, R., & Ghosh, D. (2016). A study on Improvisation in a Musical performance using Multifractal Detrended Cross Correlation Analysis. *Physica A: Statistical Mechanics and its Applications*, *462*, 67-83.
[12] Banerjee, A., Sanyal, S., Mukherjee, S., Guhathakurata, T., Sengupta, R., & Ghosh, D. (2016). How Do the Singing Styles vary over Generations in different Gharanas of Hindustani Classical Music A Comparative Non Linear Study. *arXiv preprint arXiv:1604.02250*.
[13] Ihlen, E. A. F. E. (2012). Introduction to multifractal detrended fluctuation analysis in Matlab. *Frontiers in physiology*, *3*, 141.